\documentclass[epj-spec]{svjour}
\usepackage{graphics}
\begin{document}
\title{Transverse Transport in Graphite}
\author{M. P. L\'opez-Sancho\inst{1}\fnmsep\thanks{\email{pilar@icmm.csic.es}}
\and M. A. H. Vozmediano\inst{2} \and F. Guinea\inst{1} }
\institute{Instituto de Ciencia de Materiales de Madrid-CSIC,
C/Sor Juana In\'es de la Cruz nº 3, Cantoblanco, 28046 Madrid, Spain\
\and Unidad Asociada CSIC-UC3M, Universidad Carlos III de Madrid, E-28911, Legan\'es, Madrid, Spain. }

%
%
\abstract{ Graphite is a layered material showing a strong
anisotropy. Among the unconventional properties reported by
experiments, the electronic transport along the $c$-axis, which
has direct implications in order to build graphitic devices,
remains a controversial topic. We study the influence of inelastic
scattering on the electron tunnelling between layers. In the
presence of electron electron interactions, tunnelling processes
are modified by inelastic scattering events.
} 

\maketitle
\section{Introduction}
\label{intro}

Graphite is composed of stacked layers of two-dimensional
graphene. It is known that layered materials have been object of
intensive study since they present important physics. A key
feature of many layered materials is the anisotropy exhibited by
its transport properties: while being metallic within the layers,
the transport in the $c$-axis, perpendicular to the layers, may be
coherent or incoherent and undergo a crossover with temperature
from one regime to the other, thus changing the effective
dimensionality of the system \cite{valla,esq0}. Even when coherent
electron excitations can be assumed within individual layers,
there is no consensus  about over what length and time scales the
excitations are coherent between layers \cite{valla}. Electron
correlations play as well a key role in the physics of layered
materials since correlation effects increase as dimensionality
decreases, therefore dimensionality is crucial for the electronic
properties and to choose the appropriate model to study the
system. Unusual properties are derived from the anisotropy and
periodicity along the axis perpendicular to the planes i.e. the
structure of collective excitations absent in two dimensional (2D)
and three dimensional (3D) electron gases \cite{quinn} These
unconventional properties are distinctly different from that of
the traditional Fermi liquids.

Graphite, an hexagonal layered material, presents an intraplane
hopping much larger than the interplane hybridization. Many of the
transport properties established in the past for this well known
material are being questioned at present. Recent conductivity
measurements reveal a suppression of the c--axis conductivity much
larger than what would be predicted by the band calculations of
the interlayer hopping \cite{esqui1}. Band structure calculations
are also challenged by very recent claims of observation of
quantum Hall plateaus in pure graphite \cite{esqui2}. The
unconventional transport properties of graphite such as the linear
increase with energy of the inverse lifetime \cite{Yetal96,STK01},
suggest deviations from the conventional Fermi liquid behavior,
which could be due to strong Coulomb interactions unscreened
because of the lack of states at the Fermi level
\cite{GGV94,GGV99}. The experimental isolation of single graphite
layers (graphene) \cite{Netal04} has enhanced the importance of
graphite opening new possibilities in the application field. The
properties of few of single layers of graphite have been measured
and the experiments confirm the theoretical predictions of a
physics governed by the two-dimensional Dirac equation and a
linear dispersion in the proximity of the {\bf K} points
\cite{Netal04,Netal05}

By assuming that electron correlations modify the in-plane
electron propagators \cite{TL01}, we show that even in the clean
limit, many-body effects can alter the coherent contribution to
the out of plane electron hopping. The clean limit is defined  as
that in which the length scale over which electrons remain
coherent within the layers, diverges. It is shown that, for
certain models of correlated electrons \cite{GGV94,GGV99,GGV96},
the interplane hopping between extended states can be a relevant
or an irrelevant variable, in the renormalization group (RG)
sense, depending on the strength of the coupling constant. The
interlayer coupling in graphite is still under debate. Band
structure calculations predicted an interlayer coupling much
larger than the values deduced by $c$-axis conductivity
measurements \cite{NCPG06,DRSLL04}. Furthermore, the traditional
assumption of graphite as a semimetal does not consider the
electron-electron interactions which, due to the low density of
states, remains unscreened \cite{NCPG06,GGV01}. The scheme used
here is based on the RG analysis as applied to models of
interacting electrons \cite{P92}.

Two--dimensional systems  that support low--energy excitations
which can be described by Dirac fermions have recently attired  a
renewed interest. Besides  quasiparticles of planar zero--gap
semiconductors as graphite \cite{GGV94,GGV99} there are examples
as the flux--phase of planar magnets \cite{RW01}, nodal
quasiparticles in d--wave superconductors \cite{Y01,GKR03}, or the
insulating spin density wave phase of high T$_c$ superconductors
\cite{H02}.

Long range Coulomb interactions and disorder are modelled in these
systems by means of gauge fields coupled to the Dirac fermions
\cite{VLG02}. Most of these systems show anomalous transport
properties ranging from mild departures of Fermi liquid behavior
as in pure graphite, to the total destruction of the quasiparticle
pole. Disorder can be modelled as random gauge fields coupled  to
the Dirac quasiparticles. It affects the quasiparticle Green's
function in a computable way what in turn modifies the interlayer
tunnelling.

\section{The method of calculation}
\label{method}

We analyze here the interlayer hopping in a layered material such
as graphite. In the presence of electron-electron interactions,
tunnelling processes are modified by inelastic scattering events.
The influence of inelastic scattering on electron tunnelling has
been studied, using equivalent methods, in mesoscopic devices
which show Coulomb blockade\cite{SET}, Luttinger
liquids\cite{W90,KF92,1D,SNW95}, and dirty metals\cite{RG01}. The
simplest formulation of the method replaces the excitations of the
system (such as electron-hole pairs) by a bath of harmonic
oscillators with the same excitation spectrum. This approach can
be justified rigorously in one dimension, and is always an
accurate description of the response of the system when the
coupling of the quasiparticles to each individual excitation is
weak although the net effect of the environment on the system
under study can be large\cite{CL83}. The expression for the
coupling between the electrons and the oscillators is obtained by
assuming that the oscillators describe the charge oscillations of
the system. Then the squared coupling is related, using
perturbation theory, to the charge-charge correlations in the
electron gas. This is consistent with the assumption that the
modes in the environment are weakly perturbed by their interaction
to the low energy electrons whose tunnelling properties are been
considered:

\begin{equation}
H_{int} = c^{\dag}_i c_i \sum_{\bf k}
V_i ( {\bf k} ) \hat{\rho}_{{\bf k}}
\label{int}
\end{equation}
where $c^{\dag}_i$($c_i$)  creates(destroys) an electron at site
$i$, and $\hat{\rho}_{\bf \vec{k}}$ describes the charge
fluctuations of the environment, which are to be described as a
set of harmonic modes as stated above. he interaction in
eq.(\ref{int}) is the simplest, and most common, coupling between
the tunnelling electrons and the excitations of the system, which
is spin independent. Other spin-dependent or more complicated
couplings can also be taken into account, provided that the
appropriate response function is used.

Since the excitations of the
system (electron-hole pairs, plasmons) are modelled as bosonic
modes, one can write an effective electron-boson hamiltonian of the type:
\begin{eqnarray}
 H_{e-b} &=&H_{elec} +  H_{env} + H_{int} \nonumber
  \\
&= &\sum {\bf t}_{ij} c^{\dag}_i c_j + \sum \omega_k b^{\dag}_k
b_k + \sum g_{k , i} c^{\dag}_i c_i ( b^{\dag}_k + b_k )
\label{hamil}
\end{eqnarray}
where $ H_{elec}$ describes the individual quasiparticles,
$ H_{env}$ stands for the set of harmonic oscillators
which describe the environment, and ${\cal H}_{int}$ defines
the (linear) coupling between the two.

The $b_k^{\dag}$($b_k$)
are boson creation(destruction)  operators,
the ${\bf t}_{ij}$ describe the electronic hopping processes.
The
information about the interaction between the electron in
state $i$ and the
environment is defined by the  spectral function\cite{CL83}

$$J_i ( \omega ) = \sum_k
| g_{k,i} |^2 \delta ( \omega - \omega_k ).$$

Using second order perturbation theory and eq.(\ref{int}),
we can write\cite{SET,CL83}:
\begin{equation}
J_i ( \omega ) = \sum_{\bf \vec{k}} V_i^2 ( {\bf \vec{k}} )
{\rm Im} \chi ( {\bf \vec{k}} , \omega )
\end{equation}
where $\chi ( {\bf \vec{k}} , \omega )$ is the Fourier transform
of the density-density response of the system, $\langle
\hat{\rho}_{\bf \vec{k}} ( t ) \hat{\rho}_{\bf - \vec{k}} ( 0 )
\rangle$. The electron-boson interaction leads to a Franck-Condon
factor which reduces the effective tunnelling rate. The
Franck-Condon factor depends exponentially on the coupling between
the particles and the oscillators, and when it diverges additional
self consistency requirements between the hopping amplitudes and
the oscillator frequencies included in the calculation have to be
imposed. The electron propagators acquire an anomalous time, or
energy, dependence, that can be calculated to all orders if the
state $i$ is localized, or, which is equivalent, neglecting the
hopping terms in eq.(\ref{hamil}). The present scheme can be
considered a generalization to layered systems of the approach
presented in reference\cite{KF92} for a Luttinger liquid, where
the scaling of tunnelling is expressed in terms of an effective
density of states. To second order in perturbation theory, the
electron propagators in a localized system are
\begin{eqnarray}
\langle c^{\dag}_i ( t ) c_i ( t' )
\rangle
&\sim  &\langle c^{\dag}_i ( t ) c_i ( t' ) \rangle_0 \times
\nonumber \\ & &\exp \left\{ -
\int d \omega  \left[ 1 - e^{i \omega ( t - t' )} \right]
\frac{J_{i} ( \omega )}
{ \omega^2} \right\}
\label{Green}
\end{eqnarray}
where $\langle c^{\dag}_i ( t ) c_i ( t' ) \rangle_0 \sim e^{i
\varepsilon_i ( t - t')}$ is the Green's function in the absence
of the interaction. In contrast to previous work, we  analyze
tunnelling between coherent extended states. In order to do so, we
need to generalize eq.(\ref{Green}) to this case. We then need to
know the Green's function of coherent states in the individual
layers, $G( {\bf \vec{k}}, w)$, including the correction due to
the interaction to the environment. We firstly  assume that
eq.(\ref{Green}) also holds in a system with extended states. For
a standard metallic system, we must insert $\langle c^{\dag}_i ( t
) c_i ( t' ) \rangle_0 \sim 1 / ( t - t' )$ in eq.(\ref{Green}).
It can be shown that this approximation is exact at short times,
$W \ll ( t - t' )^{-1} \ll \Lambda$, where $W$ is an energy scale
related to the  dynamics of the electrons, and $\Lambda$ is the
upper cutoff in the spectrum of the environment. This expression
can be generalized, taking into account the spatial structure of
the coupling to:
\begin{eqnarray}
\langle \Psi^{{\dag}}_i ( t ) \Psi_j ( t' )
\rangle &
\sim &\frac{1}{t - t'} \times \nonumber \\ &\times &\exp \left\{
\int
 \int_{ \Omega} \int_{ \Omega }
 d^2 {\bf \vec{r}} d^2 {\bf \vec{r}'}
 d^2 {\bf \vec{k}} e^{i {\bf \vec{k}}
 ( {\bf \vec{r} - \vec{r}'} )} \right. \nonumber \\ & & \left.
 \int d \omega \left[ 1 - e^{i \omega ( t - t' )} \right]
 \frac{V_{eff} ( {\bf \vec{k}} , \omega )}
 { \omega^2} \right\}
 \label{propagator}
 \end{eqnarray}
 where $\Omega$ is the region of overlap of the wavefunctions
 $\Psi_i ( {\bf \vec{r}} )$ and
 $\Psi_j ( {\bf \vec{r}} )$.
 This expression, which can be seen as the exponential of the leading
 frequency dependent self--energy correction to the electron propagator,
 has been extensively used in studies of tunnelling in zero dimensional
 systems (single electron transistors) which show Coulomb
 blockade\cite{SET}, one dimensional conductors\cite{SNW95}, and
 disordered systems in arbitrary dimensions\cite{RG01}.

 The effective interaction can, in turn, be written in terms
 of the response function as:
 \begin{equation}
 V_{eff} ({\vec {\bf k}} , \omega ) = V^2 ( {\vec {\bf k}} )
 {\rm Im} \chi ({\vec {\bf k}} , \omega )
 \label{veff}
 \end{equation}

 The time dependence in eq. (\ref{propagator}) is determined
 by
 the low energy limit of the response function.
 In a Fermi liquid, we have
 \begin{equation}
 {\rm Im} \chi {\vec ({\bf k}} , \omega ) \approx \alpha ( {\vec{\bf k}} )
 \vert \omega \vert \;\;\;  \omega \ll E_F\      \;,
 \end{equation}
 where $E_F$ is the Fermi energy.
 We then get:
 \begin{equation}
 \lim_{(t-t') \rightarrow \infty} \langle \Psi^{{\dag}} ( t ) \Psi ( t' )
 \rangle
 \sim \frac{1}{( t - t' )^{1 + {\alpha}}}
 \label{green}
 \end{equation}
 where
 \begin{equation}
 {\alpha} = \int_{{\bf \vec{k}}}\int_{\Omega} \int_{\Omega}
 d^2 {\bf \vec{r}} d^2 {\bf \vec{r}}'
 d^2 {\bf \vec{k}} e^{i {\bf \vec{k}}
 ( {\bf \vec{r} - \vec{r}'} )} V^2( {\bf \vec{k}} ) \alpha ( {\bf \vec{k}} )
 \label{alpha}
 \end{equation}
the parameter ${\alpha}$ gives the correction to the scaling
properties of the Green's functions. Integration in ${\bf
\vec{k}}$ is restricted to $ | {\bf \vec{k}} | \ll L^{-1}$ where
$L$ is the scale of the region where the tunnelling process takes
place. The value of $L$ is limited  by the length over which the
phase of the electronic wavefunctions within the layers is well
defined.

We can now use eq.(\ref{green}) to analyze the interlayer
tunnelling by applying renormalization group methods. The simplest
case where this procedure has been used is for the problem of an
electron tunnelling between two states, $i$ and $j$, which has
been intensively studied\cite{Letal87,W93}.  We integrate out the
high energy bosons, with energies $\Lambda - d \Lambda \le
\omega_k \le \Lambda$ and rescaled hopping terms are defined. As
mentioned earlier, eq.(\ref{green}) is valid for this range of
energies. The renormalization of the hoppings is such that the
properties of the effective Hamiltonian at energies $\omega \ll
\Lambda$ remain invariant. If the hoppings ${\bf t}_{ij}$ are
small, any physical quantity which depends on them can be
expanded, using time dependent perturbation theory, in powers of:
\begin{equation}
{\bf t}_{ij}^2 \langle c_i^{\dag} ( t ) c_j ( t ) c_j^{\dag} ( t' ) c_i ( t' )
\rangle \approx {\bf t}_{ij}^2 \langle c_i^{\dag} ( t ) c_i ( t' ) \rangle
\langle c_j ( t ) c_j^{\dag} ( t' ) \rangle
\label{perturbation}
\end{equation}
The integration of the high energy modes implies that the terms
in eq.(\ref{perturbation}) are restricted to $t \le \Lambda^{-1}$,
or, alternatively, the time  unit have to be rescaled\cite{C81},
$\tau' = \tau e^{d\Lambda / \Lambda }$, where $\tau \sim
\Lambda^{-1}$.
Using eq.(\ref{green}), the condition of keeping the
perturbation expansion in powers of the terms in eq.(\ref{perturbation})
invariant implies that:
\begin{equation}
{\bf t}_{ij}^2 \rightarrow {\bf t}_{ij}^2  e^{
\frac{d \Lambda}{\Lambda} \left( 2 + 2 \alpha \right)}
\end{equation}
which can also be used to define the scaling dimension of
the hopping terms. Finally,
\begin{equation}
\frac{\partial ( {\bf t}_{ij} / \Lambda )}{\partial l} =
- \alpha \frac{ {\bf t}_{ij} }{\Lambda}
\label{renor}
\end{equation}
where $l = \log ( \Lambda_0 / \Lambda )$, and $\Lambda_0$
is the initial value of the cutoff.

This approach has been successfully used to describe inelastic
tunnelling in different situations
in\cite{TL01,SET,W90,KF92,1D,SNW95,RG01}.

The analysis which leads to eq.(\ref{renor}) can be generalized to
study hopping between extended states, provided that we
can estimate the long time behavior of the Green's
function,
\begin{equation}
G ( {\bf \vec{k}} , t - t' )
= \langle c^{\dag}_{\bf \vec{k}} ( t ) c_{\bf \vec{k}} ( t' )
\rangle .
\label{gtk}
\end{equation}

We assume that, in a translationally invariant system,
there is no dependence on the position of the local orbital, $i$.
This result implies that the frequency dependence of
the Green's function, in a
continuum description, can be written as:
\begin{equation}
\lim_{|{\bf \vec{r}} - {\bf \vec{r}}' | \rightarrow 0}
G ( {\bf \vec{r}} - {\bf \vec{r}}', \omega )
\propto | \omega |^{ \alpha}
\label{green_w}
\end{equation}
Equation(\ref{gtk}) is related to  eq.(\ref{green_w}), by:
\begin{equation}
\lim_{|{\bf \vec{r}} - {\bf \vec{r}}' | \rightarrow 0}
G ( {\bf \vec{r}} - {\bf \vec{r}}', \omega ) =
\int d^D {\bf \vec{k}} G ( {\bf \vec{k}} , \omega )
\label{integral}
\end{equation}
where $D$ is the spatial dimension.
In the cases discussed below, the interaction is instantaneous in time,
and the non interacting Green's function can be written as:
\begin{equation}
G_0 ( {\bf \vec{k}} , \omega ) \propto \frac{1}{\omega}
{\cal F} \left( \frac{k_i^z}{\omega} \right)
\label{scaling_g0}
\end{equation}
where $z=1,2$ depending on the dispersion relation of the given model.
In the following, we
assume that the interacting Green's function has the
same scaling properties, with the factor $\omega^{-1}$
replaced by $\omega^{-\delta}$ in eq.(\ref{scaling_g0}), where $\delta$
depends on the interactions.
This can be shown to be correct
in perturbation theory to all orders, in the models
studied below, which describe the physics in the proximity
of critical points, because the corrections depend logarithmically
on $\omega$ (it is a well known fact for the
Luttinger liquid). Then, using eqs.
(\ref{green_w}), (\ref{integral})
and (\ref{scaling_g0}), we obtain:
\begin{equation}
G ( {\bf \vec{k}} , \omega ) \propto
| \omega |^{\alpha - D/z} {\cal F} \left( \frac{k_i^z}{\omega} \right)
\label{green_k}
\end{equation}
and $ {\cal F} ( u )$ is finite.
Thus, from the knowledge of
the real space Green's function, using
eq.(\ref{Green}),  we obtain $\alpha$, which, in turn, determines
the exponent $\alpha + D / z $ which characterizes
$G ( {\bf \vec{k}} , \omega )$. Generically, we can write:
\begin{equation}
G_{loc,ext} ( \omega ) \sim | \omega |^{\delta_{loc,ext}}
\label{scaling_green}
\end{equation}
where the subindices $loc , ext$ stand for localized and extended
wavefunctions. In terms of these exponents, we can generalize
eq.(\ref{renor}) to tunnelling between general states to:
\begin{equation}
\frac{\partial ( {\bf t}_{ij}^{loc,ext} / \Lambda )}{\partial l} =
- \delta_{loc,ext} \frac{ {\bf t}_{ij} }{\Lambda}
\label{renor2}
\end{equation}
Before proceeding to calculations of $\delta_{loc}$
and $\delta_{ext}$ for various models,
it is interesting to note that, in general,
the response function of an electron gas in dimension $D > 1$ behaves as
$$\lim_{\omega \rightarrow 0 , | {\bf \vec{k}} | \rightarrow 0}
\chi( {\bf \vec{k}} , \omega )
 \sim | \omega | / | {\bf \vec{k}} | ,$$
 so that, from
 eq.(\ref{alpha}),
 $$\lim_{L \rightarrow
 \infty} \alpha \sim L^{(1-D)}.$$
 Thus, for $D > 1$, the contribution of the inelastic processes
 to the renormalization of the tunnelling vanishes for delocalized
 states, $L \rightarrow \infty$.

It is easy to show that, in an isotropic Fermi liquid in D dimensions,
$\lim_{L \rightarrow \infty}
{\alpha} \propto L^{1 - D}$, where $L$ is the linear
dimension of the (localized) electronic wavefunctions $\Psi
( {\bf \vec{r}} )$. This result is due to
the dependence on ${\bf \vec{k}}$ of the response
functions which goes as
$${\rm Im} \chi ( {\bf \vec{k}} ,
\omega ) \sim\frac{ | \omega | }{  k_F^{D-1} | {\bf \vec{k}} |
}.$$ Thus, for $D > 1$, we recover coherent tunnelling in the
limit of delocalized wavefunctions.

In one dimension, one can use the non interacting
expression for ${\rm Im} \chi_0 ( {\bf \vec{k}} ,
\omega )$, to obtain:
\begin{eqnarray}
& &\int
d^2 {\bf \vec{r}} d^2 {\bf \vec{r}'}
d^2 {\bf \vec{k}} e^{i {\bf \vec{k}}
( {\bf \vec{r} - \vec{r}'} )}
\int d \omega \left[ 1 - e^{i \omega ( t - t' )} \right]
\frac{V_{eff} ( {\bf \vec{k}} , \omega )}
{ \omega^2} \nonumber \\
&\propto &\left( \frac{U}{E_F} \right)^2 \times \left\{
\begin{array}{lr} 0 &t - t' \ll L / v_F \\
\log [ v_F ( t - t' ) / L ] &t - t' \gg L / v_F \end{array}
\right.
\label{1D}
\end{eqnarray}
where we have assumed a smooth short range interaction,
parametrized by $U$. Hence, the Green's functions have a non
trivial power dependence on time, even in the $L \rightarrow
\infty$ limit, in agreement with well known results for Luttinger
liquids\cite{SNW95}. In order to obtain the energy dependence of
the effective tunnelling between ${\bf \vec{k}}$ states near the
Fermi surface, one needs to perform an additional integration over
$d {\bf \vec{r}}$. In general, near a scale invariant fixed point,
$\omega \propto | {\bf \vec{k}} |^z$, and for a 1D conductor one
knows that $z=1$. Hence,
$${\rm Im}\; G ( \omega , k_F )
\propto \omega^{-z + {\alpha}} \sim \omega^{-1 + {\alpha}}.$$ The
flow of the hopping terms under a renormalization group scaling of
the cutoff is \cite{1D}:
\begin{equation}
\frac{\partial ( {\bf t} / \Lambda )}{\partial l} =
\left\{ \begin{array}{lr} - {\alpha} &{\rm localized \, \, \, hopping} \\
1 - {\alpha} &{\rm extended \, \, \, hopping} \end{array} \right.
\label{scaling}
\end{equation}
where ${\bf t}$ denotes a hopping term, between localized or extended states.
In the latter case,
the hopping becomes
an irrelevant variable\cite{W90,CGP06,VLG03} for ${\alpha} > 1$.

\section{Graphite layers}
\label{Graphite}

The simplest two dimensional model for interacting electrons where
it can be rigourously shown that the couplings acquire logarithmic
corrections in perturbation theory is a system of Dirac fermions
($\epsilon_k = v_F | {\bf \vec{k} |}$), with Coulomb, $1 / |
{\bf \vec{r} - \vec{r}'} |$, interaction. This model can be used
to describe isolated graphene planes\cite{GGV94,GGV99}, formed by
carbon atoms arranged i the honeycomb lattice, and can
help to understand the anomalous behavior of graphite observed
in some experiments\cite{Yetal96,STK01}.

In order to apply the procedure outlined in the previous section, one
needs the Fourier transform of the interaction,
$$V_{eff}({\bf \vec{k}})=e^2 / ( \epsilon_0
| {\bf \vec{k}} |),$$
where $e$ is the electronic charge, and
$\epsilon_0$ is the dielectric constant, and the susceptibility
of the electron gas. For a single graphene plane, this quantity
has been computed in \cite{GGV94} and is:
\begin{equation}
\chi_0 ({\bf \vec{k}}, \omega ) = \frac{1}{8} \frac{ | {\bf
\vec{k}}|^2} {\sqrt{v_F^2 |{\bf \vec{k}}|^2 - \omega^2 }}
\label{chigraphite}
\end{equation}
These expressions need to be inserted
in equations (\ref{veff}) and (\ref{propagator}).
Alternatively, we can use the RPA, and include the effects
of interplane screening:

\begin{equation}
\begin{array}{ll}
\chi_{RPA} ({\vec{\bf k}}, \omega  ) &= \\
& \frac{\sinh ( | {\bf \vec{k}} | d )}
{\sqrt{ \left[ \cosh ( | {\vec{\bf k}} | d ) + \frac{2 \pi e^2}
{ \epsilon_0 | {\bf k} |}
\sinh ( | {\vec{\bf k}} | d )
\chi_0 ( {\vec{\bf k}} , \omega ) \right]^2 - 1 }} \end{array}
\label{RPA}
\end{equation}
where $d$ is the interplane spacing. The imaginary part, ${\rm Im} \chi_0
( {\bf \vec{k}} , \omega )$, is different from zero if $\omega >
v_F | {\bf \vec{k}} |$.

For simplicity, we consider the expression in
eq.(\ref{chigraphite}), as it allows us to obtain analytical
results. We cut off the spatial integrals at a scale, $L$, of the
order of the electronic wavefunctions involved in the tunnelling.
Performing the same computation as in the case of D=1 we obtain an
expression similar to that in eq.(\ref{1D}) except that the
prefactor $( U / E_F )^2$ is replaced by the squared of  effective
coupling constant of the model, $e^4 / ( \epsilon_0 v_F )^2$.
Thus, also  in the graphene model, the propagators acquire an
anomalous dimension what was advocated  in \cite{GGV94} as
pointing out to a departure of the model from Fermi liquid
behavior. As in 1D, the value of the exponent $z$ which relates
length and time scales is $z=1$. The scaling of the hoppings now
are:
\begin{equation}
\Lambda \frac{\partial ( {\bf t} / \Lambda )}{\partial \Lambda} =
\left\{ \begin{array}{lr} -
1 - {\alpha} &{\rm localized \, \, \, hopping} \\
1 - {\alpha} &{\rm extended \, \, \, hopping} \end{array} \right.
\label{scaling_graphite}
\end{equation}
The extra constant in the first equation with respect to eq. (\ref{scaling})
reflects the vanishing of the density of states at the Fermi level for
two dimensional electrons with a Dirac dispersion relation.

In graphite, the dimensionless coupling constant,  $e^2 / v_F$, is
of order unity. Under renormalization, it flows towards
zero\cite{GGV94}. Thus, despite the departure from Fermi liquid
behavior, interplane tunnelling is a relevant variable and a
coherent out-of-plane transport in pure clean graphene samples
should be observed\cite{VLG02}.
\subsection{Influence of disorder.}
\label{Disorder}
In previous section, the calculation has been done in the clean
limit. This picture can change in the presence of disorder that is
known to change the anomalous dimension of the fields. Disorder
can be incorporated in the Fermi liquid two dimensional systems in
the clean limit. In the renormalization group scheme disorder can
be included by the introduction of random gauge fields
\cite{CHA06,SGV05}. In fact, this is a standard procedure in the
study of the states described by the two-dimensional Dirac
equation associated to random lattices or to integer quantum Hall
transitions and, usually, the density of states at low energies is
increased. Topological disorder, as pentagons and heptagons in the
hexagonal lattice, have been observed in graphite and carbon
nanotubes samples. A random distribution of topological defects
can be described by a non-Abelian random gauge in the
calculation\cite{SGV05} and this features will influence the
interlayer transport properties\cite{VLG03}. Disorder in general,
will affect the electronic properties in different ways.
Ferromagnetic behavior of graphite samples, enhanced by proton
irradiation, has been reported \cite{ESHSHB03}. The formation of
vacancies, cracks or voids in the graphite lattice, due to the
proton bombardment, favors the magnetic order in the sample due to
the localized states which arise at the borders of the defects
\cite{VLSG05}, where the change of coordination of the carbon
atoms gives rise to  the existence of unpaired spins. The
incompletely screened electron-electron interactions will induce
the ferromagnetic correlation between these unpaired spins
\cite{VLSG05,PGC05} in the layers. The influence of the long-range
Coulomb correlation in the interlayer coupling remains to be
completely understood. Recently The possibility of making atomic
scale devices based on graphene layers has been suggested by the
experiment. The control of the electronic structure of a bilayer
graphene has been reported\cite{OBSHR06}. The gap between the
valence and conduction bands has been controlled by selectively
adjusting the carrier concentration of each layer, to change the
Coulomb potential, at higher electron density the overlap between
orbitals of adjacent layers increases, thus increasing the
interlayer coupling. The interlayer interaction varies then as a
function of the electron concentration in the layers.

The possibility of controlling as well the spin transport in graphene and in CNTs, is
being currently investigated. From the theoretical point of view, the inclusion
of the spin-orbit interaction in the Hamiltonian\cite{CLM04,KM05,HGB06} may explain
some results of long spin correlation length in CNTs contacted to ferromagnetic leads.
.
\section{Conclusions}

The potential applications of graphite sheets together with those
of CNTs have caused a renewed interest in carbon-based materials.
In order to understand their behavior under external probes, it is
important to know the dielectric response, which is related to the
collective excitations of the electrons which combine in-plane and
inter-plane interactions. Therefore the exact nature of the
interlayer coupling is needed. We have analyzed the relation
between the interlayer tunnelling and the inelastic processes in
the layers, in the clean limit at low energy. Our results suggest
that, when perturbation theory for the in--plane interactions
leads to logarithmic divergences, the out-of-plane tunnelling
acquires a non trivial energy dependence. This anomalous scaling
of the interlayer hopping can make it irrelevant, at low energies,
if the in--plane interactions are sufficiently strong\cite{VLG03}.
A relation between the nature of in-plane excitations and the
perpendicular transport has been reported from ARPES and transport
measurements in strongly correlated layered materials: when
coherent quasiparticles form in the planes, the perpendicular
transport becomes metallic\cite{valla}. In graphite, we find that,
despite the deviation from Fermi liquid behavior, coherent
out-of-plane transport can occur. This result makes contact with
experiments which suggest that the temperature dependence of the
out-of-plane resistivity is directly correlated to that of
in-plane resistivity in graphite\cite{KEK05}. As well, the
monotonic increase of the interlayer interaction as a function of
electron concentration in both layers recently observed in  a
bilayer graphene\cite{OBSHR06} agrees with our results.


{\it Acknowledgements}.
Funding from MEC (Spain) through grant FIS2005-05478-C02-01
and from the European Union Contract No. 12881 (NEST) is
acknowledged.

%

%
%

\end{document}